\documentclass{article}

\usepackage{PRIMEarxiv}
\usepackage[utf8]{inputenc} % allow utf-8 input
\usepackage[T1]{fontenc}    % use 8-bit T1 fonts
\usepackage{hyperref}       % hyperlinks
\usepackage{url}            % simple URL typesetting
\usepackage{booktabs}       % professional-quality tables
\usepackage{amsfonts}       % blackboard math symbols
\usepackage{nicefrac}       % compact symbols for 1/2, etc.
\usepackage{microtype}      % microtypography
\usepackage{lipsum}
\usepackage{fancyhdr}       % header
\usepackage{graphicx}       % graphics
\graphicspath{{media/}}     % organize your images and other figures under media/ folder

\title{Loop dynamics of a fully discrete short pulse
	equation
}

\author{
H.~ Sarfraz\thanks{hira\_phys@yahoo.com (Corresponding Author)}, U.~Saleem\thanks{usman\_physics@yahoo.com and usaleem.physics@pu.edu.pk} and Y.~Hanif \thanks{yasir\_pmc@yahoo.com}\\
	Department of Physics \\
	University of the Punjab, Quaid-e-Azam Campus \\
	Lahore-54590, Pakistan\\
}

\begin{document}
	\maketitle

	\begin{abstract}
		In this article, a fully discrete short pulse (SP) equation is
		presented as an integrability condition of a linear system of
		difference equations (also known as discrete Lax pair).
		Additionally, two semi-discrete versions of the SP equation have
		also been obtained from fully discrete SP equation under continuum
		limits. Darboux transformation is employed to compute multi-soliton
		solutions of fully discrete and semi-discrete SP equations. Explicit
		expressions of first and second nontrivial soliton solutions are
		computed. We also derived explicit expression of breather solution
		for fully discrete SP equation. The dynamics of single loop soliton
		and interaction mechanism of loop-loop and loop-antiloop solutions
		has been explored and illustrated.
		
	\end{abstract}

	% keywords can be removed
	
	\section{Introduction}
Nonlinear Schr\"{o}dinger equation can best describe
the propagation of slowly varying envelopes in nonlinear dispersive
media. However, when the width of the optical pulse further reduces
to the order of femtoseconds, then NLS equation cannot be derived
for such ultra-short optical pulses \cite{rothenberg1992space}. So
in order to describe the propagation of ultra-short optical signals,
slowly varying optical models needs to be modified. Sh\"{a}fer and
Wayne introduced the short pulse (SP) equation
\cite{schafer2004propagation}
\begin{equation}
	q_{xt}=q+\frac{1}{6}\left(q^3\right)_{xx}, \label{sp01}
\end{equation}
to describe the processes including ultra-short optical pulses. Here
real-valued function $q=q(x,t)$ represents the magnitude of electric
field and the subscripts $x,$ $t$ indicate usual space and time
derivatives. Short pulse equation first emerged during the study of
pseudospherical surfaces \cite{rabelo1989equations}. SP equation is
completely integrable equation and its integrability has been proven
by many means, for example, existence of associated isospectral
problem also known as Wadati-Konno-Ichikawa (WKI) scheme
\cite{sakovich2005short}, bi-Hamiltonian structure
\cite{brunelli2006bi}, infinite set of conservation laws
\cite{zhang2015conservation}, soliton solutions
\cite{matsuno2008periodic, matsuno2007multiloop, parkes2008some,
	tb2007two, sakovich2006solitary, brunelli2005short}. By using a
suitable hodograph transformation, the SP equation can be
transformed into some well-known integrable equations, such as
sine-Gordon equation \cite{matsuno2007multiloop,
	matsuno2008periodic}.

The general integrable short pulse (SP) equation  as presented in
\cite{zhaqilao2017} is given by
\begin{equation}
	q_{xt}=q+\frac{1}{2}(qrq_{x})_{x},\quad \quad \quad \quad \quad \quad r_{xt}=r+%
	\frac{1}{2}(rqr_{x})_{x},  \label{nsp02}
\end{equation}
where $q(x,t)$ and $r(x,t)$ are the dynamical variables that
represent the magnitude of the electric field. The coupled system
(\ref{nsp02}) can be written as the consistency condition of the
following linear eigenvalue problem of Wadati-Konno-Ichikawa (WKI)
type
\begin{eqnarray}
	\Psi _{x}&\equiv & U\Psi=\lambda \left(
	\begin{array}{cc}
		1 & 0 \\
		0 & -1%
	\end{array}%
	\right)\Psi +\lambda \left(
	\begin{array}{cc}
		0 & q_{x} \\
		r_{x} & 0%
	\end{array}%
	\right)\Psi,\nonumber \\
	\Psi _{t}&\equiv &V\Psi=\frac{\lambda }{2}\left(
	\begin{array}{cc}
		qr & qrq_{x} \\
		rqr_{x} & -qr%
	\end{array}%
	\right)\Psi +\frac{1}{2}\left(
	\begin{array}{cc}
		0 & -q \\
		r & 0%
	\end{array}%
	\right)\Psi +\frac{1}{4\lambda }\left(
	\begin{array}{cc}
		1 & 0 \\
		0 & -1%
	\end{array}%
	\right)\Psi,  \label{nsp03}
\end{eqnarray}%
with $\Psi=\Psi(x,t;\lambda)$. The consistency condition of spectral
problem (\ref{nsp03}) becomes a
zero-curvature condition, that is, $U_{t}-V_{x}+UV-VU=0$ which leads the SP equation (\ref%
{nsp02}).

Hodograph transformation is exclusively used to solve a nonlinear
systems with a switched role of dependent and independent variables.
This transformation is more or less similar to reciprocal
transformation except for a difference that is the reciprocal
transformation can put a system into conservative form. The
reciprocal transformation was presented by Kingston and Roger in
1982 \cite{kingston1982reciprocal}. Lets define the hodograph
transformation $\left( x,t\right) \rightarrow \left( y,\tau\right) $
by the means of following transformations relating the old variables
$\left( x,t\right) $ to the new ones $\left( y,\tau\right) $ as
\begin{equation}
	dy =\omega dx+\frac{1}{2}qr\omega dt, \quad \quad \quad \quad
	\quad\quad d \tau =dt, \label{HDT 1}
\end{equation}%
where $\omega=\sqrt{1+q_xr_x}$, also%
\begin{equation}
	\frac{\partial }{\partial x}=\omega \frac{\partial }{\partial
		y},\quad \quad \quad \quad \quad \quad \quad \quad\frac{\partial }{\partial t}=\frac{1}{2}qr\omega \frac{\partial }{%
		\partial y}+\frac{\partial }{\partial \tau}. \label{HDT 2}
\end{equation}%
Now under the transformation (\ref{HDT 1})-(\ref{HDT 2})
the equations (\ref{nsp02}) reduce to following system of equations%
\begin{equation}
	x_{y\tau }=-\frac{1}{2}\left( qr\right) _{y},\quad \quad \quad \quad%
	q_{y\tau }=qx_{y},\quad \quad \quad \quad r_{y\tau }=rx_{y}.
	\label{nsp07}
\end{equation}
Under the reduction $r=-q$ system of equations (\ref{nsp07})
becomes
\begin{equation}
	x_{y\tau }=\frac{1}{2}\left( q^2 \right)_{y},\quad \quad \quad \quad%
	q_{y\tau }=qx_{y}. \label{nsp07a}
\end{equation}
The integrable system of equations (\ref{nsp07}) can be written as
the
consistency condition of the following WKI scheme%
\begin{equation}
	\Psi _{y}=\lambda \left(
	\begin{array}{cc}
		x_{y} & q_{y} \\
		r_{y} & -x_{y}%
	\end{array}%
	\right)\Psi ,\quad \quad \quad \quad \quad \quad \quad \quad \quad
	\quad \quad \quad \quad \Psi _{\tau }=\frac{1}{2}\left(
	\begin{array}{cc}
		\frac{1}{2\lambda } & -q \\
		r & -\frac{1}{2\lambda }%
	\end{array}%
	\right)\Psi . \label{nsp08}
\end{equation}%

During last two decades, discretizations of linear and nonlinear
differential equations (ordinary and partial) have been studied
extensively. Generally speaking, many physical phenomenon are
modeled as differential-difference or difference-difference
equations. Discretization of nonlinear integrable models plays an
important role in various fields of science, e.g., in biological
sciences, nonlinear optical communications, other nonlinear systems,
optical fiber communication, quantum mechanics, field theories, etc
\cite{kosmann2004discrete, davydov1973theory, kenkre1986self,
	papanicolaou1987complete}. Such systems comprise of inherent
nonlocality.

The construction of discrete analogue of any particular integrable
system has a remarkable history. First of all, Ablowitz and Ladik
presented a linear system of difference-differential equations also
known as Ablowitz-Ladik scheme \cite{ablowitz1975nonlinear,
	ablowitz1977solution}. Later on, Hirota studied discrete integrable
analogues of some famous PDEs through Hirota bilinearization method
\cite{hirota1977nonlinear1, hirota1977nonlinear2,
	hirota1977nonlinear3, hirota1978nonlinear4, hirota1979nonlinear5}.
Integrable discretization of SP equation and its multi-component
generalizations have been presented in the articles such as
\cite{feng2010integrable, feng2015integrable, feng2011discrete,
	feng2015integrable1} by using Hirota's direct method and
self-adaptive moving mesh schemes.

In recent times, different semi-discrete versions of SP equations
have been extensively investigated. However, fully discrete SP
equation has never been given much attention. In this paper, we
would present fully discrete SP equation by considering both
variables (time and space) as discrete integral variables. In
section $2$, a generalized version of fully discrete SP equation
will be presented as consistency condition of a discrete linear
system. Moreover, under various continuum limits, fully discrete
system will be deduced to two semi-discrete versions of SP equation
(discrete in time and space separately). Section $3$ will be devoted
to a brief review of discrete Darboux transformation to construct
discrete multi-soliton solutions through the covariance of discrete
Lax pair. In section $4$, explicit expressions of first two
nontrivial solutions will be computed. Fascinating interactions of
loop-loop and loop-antiloop will be depicted. Furthermore, discrete
breather for a particular choice of spectral parameters will be
obtained too. The article will end up with a concise summary.
	
	\section{Discretization}
	In this section, we will present the fully discrete, time
	discretized and space discretized versions of SP equation.
	Additionally, the results obtained here can be reducible to already
	known versions of SP equation under continuum limits.
	\subsection{Fully discrete SP equation}
	A generic fully discrete SP equation is given by
	\scriptsize
	\begin{eqnarray}
		&&\Delta_n \Delta_m x_{n,m}+\frac{b}{4}\left(\Delta_n
		x_{n,m+1}\Delta_m x_{n,m}-\Delta_m x_{n+1,m}\Delta_n
		x_{n,m}\right)+\frac{b}{4} \nonumber \\
		&&\left(\nabla_m \Delta_n q_{n,m}\nabla_n r_{n,m}+\nabla_m \Delta_n
		r_{n,m}\nabla_m q_{n,m}+\Delta_n r_{n,m}\nabla_m \Delta_n
		q_{n,m}+\Delta_n q_{n,m}\nabla_m \Delta_n r_{n,m}\right)=0,\nonumber
		\\
		&& \label{DSP1}\\
		&&\Delta_n \Delta_m q_{n,m}-\frac{b}{4}\left(\Delta_n \Delta_m
		x_{n,m}\nabla_m q_{n,m}+\Delta_n \Delta_m x_{n,m}\Delta_n
		q_{n,m}+\Delta_n x_{n,m}\nabla_m \nabla_n q_{n,m}-\Delta_n
		x_{n,m}\Delta_n \Delta_m q_{n,m}\right)=0,\nonumber \\
		&& \label{DSP2} \\
		&&\Delta_n \Delta_m r_{n,m}-\frac{b}{4}\left(\Delta_n \Delta_m
		x_{n,m}\nabla_m r_{n,m}+\Delta_n \Delta_m x_{n,m}\Delta_n
		r_{n,m}+\Delta_n x_{n,m}\nabla_m \nabla_n r_{n,m}-\Delta_n
		x_{n,m}\Delta_n \Delta_m r_{n,m}\right)=0.\nonumber \\
		&& \label{DSP3}
	\end{eqnarray}
	\normalsize
	Here difference operator ($\Delta_{a}$) and forward shift operator
	($\nabla_a$) are defined on an arbitrary discrete-valued function
	$l_{n,m}$ as
	\begin{eqnarray*}
		\Delta_n l_{n,m}=l_{n+1,m}-l_{n,m}, \quad \Delta_m
		l_{n,m}=l_{n,m+1}-l_{n,m}, \\  \nabla_n
		l_{n,m}=l_{n+1,m}+l_{n,m}, \quad  \nabla_m
		l_{n,m}=l_{n,m+1}+l_{n,m}.
	\end{eqnarray*}
	The generalized fully discrete SP equation (\ref{DSP1})-(\ref{DSP3})
	appears as compatibility condition of the following linear system of
	difference equations
	\begin{eqnarray}
		L_n\Psi_{n,m}&\equiv & \Psi_{n+1,m}=\left(I_2+\lambda Q_{n,m}\right)\Psi_{n,m}\nonumber \\
		&=& \left(\begin{array}{cc} 1+\lambda \Delta_n x_{n,m} &
			\lambda \Delta_n q_{n,m} \\
			\lambda \Delta_n r_{n,m} & 1-\lambda \Delta_n x_{n,m}
		\end{array}
		\right)\Psi_{n,m}, \nonumber \\
		L_m\Psi_{n,m}&\equiv &\Psi_{n,
			m+1}=\left(I_2+\frac{b}{4\lambda}\sigma_3+\frac{b}{4}R_{n,m}\right)\Psi_{n,m}\nonumber
		\\
		&=& \left(\begin{array}{cc}
			1+\frac{b}{4\lambda}+\frac{b}{4}\Delta_{m} x_{n,m} &
			-\frac{b}{4}\nabla_m q_{n,m}\\
			\frac{b}{4}\nabla_m r_{n,m} &
			1-\frac{b}{4\lambda}+\frac{b}{4}\Delta_{m}
			x_{n,m}\end{array}\right)\Psi_{n,m}. \nonumber \\
		&&\label{LP}
	\end{eqnarray}
	where
	\begin{equation}
		Q_{n,m}=\left(\begin{array}{cc}  \Delta_n x_{n,m} &
			\Delta_n q_{n,m} \\
			\Delta_n r_{n,m} & -\Delta_n x_{n,m}
		\end{array}
		\right), \quad \quad \quad R_{n,m}= \left(\begin{array}{cc} \Delta_m
			x_{n,m} &
			-\nabla_m q_{n,m} \\
			\nabla_m r_{n,m} & \Delta_m x_{n,m}
		\end{array}
		\right).
	\end{equation}
	Here $n$ and $m$ are discrete variables and $\lambda$ is a real or
	complex-valued spectral parameter. The integrability condition of
	(\ref{LP}) i.e., $L_n \Psi_{n,m}=L_m\Psi_{n,m}$ will equivalently
	yield discrete SP equation (\ref{DSP1})-(\ref{DSP3}). Therefore, the
	linear system of difference-difference equations (\ref{LP}) is also
	named as fully discrete Lax pair. Under the continuum limit $a
	\rightarrow 0$ and $b \rightarrow 0$ discrete SP equation
	(\ref{DSP1})-(\ref{DSP3}) and its associated Lax pair (\ref{LP})
	will reduce into continuous SP equation and its associated linear
	system (\ref{nsp07}) and (\ref{nsp08}) respectively. Under the
	reduction $r_{n,m}=-q_{n,m}$, system (\ref{DSP1})-(\ref{DSP3})
	reduces to
	\begin{eqnarray}
		&&\Delta_n \Delta_m x_{n,m}+\frac{b}{4}\left(\Delta_n
		x_{n,m+1}\Delta_m x_{n,m}-\Delta_m x_{n+1,m}\Delta_n
		x_{n,m}\right)- \nonumber \\
		&&\frac{b}{4}\left(\nabla_m \Delta_n q_{n,m}\nabla_n q_{n,m}+\nabla_m \Delta_n
		q_{n,m}\nabla_m q_{n,m}+2\Delta_n q_{n,m}\nabla_m \Delta_n
		q_{n,m}\right)=0,\nonumber
		\\
		&& \label{DSP1a}\\
		&&\Delta_n \Delta_m q_{n,m}-\frac{b}{4}\left(\Delta_n \Delta_m
		x_{n,m}\nabla_m q_{n,m}+\right.\nonumber\\&&\left.\Delta_n \Delta_m x_{n,m}\Delta_n
		q_{n,m}+\Delta_n x_{n,m}\nabla_m \nabla_n q_{n,m}-\Delta_n
		x_{n,m}\Delta_n \Delta_m q_{n,m}\right)=0.\nonumber \\
		\label{DSP2b}
	\end{eqnarray}
	Equations (\ref{DSP1a})-(\ref{DSP2b}) represent a fully discrete
	parametric (transformed) version of SP equation (\ref{nsp07a}).
	\subsection{Semi-discrete SP equation (discrete in time)}
	Under the continuum limit $a \rightarrow 0$ and $n \rightarrow
	\infty$, the discrete linear system (\ref{LP}) will reduce to
	following pair of differential-difference equations:
	\begin{eqnarray}
		\frac{d}{dy}\Psi_m&=& \left( \begin{array}{cc} \lambda\frac{d}{dy}
			x_m & \lambda\frac{d}{dy} q_m \\
			\lambda\frac{d}{dy} r_m & -\lambda\frac{d}{dy} x_m \end{array}
		\right)\Psi_m, \nonumber \\
		L_m \Psi_m\equiv \Psi_{m+1}&=&\left(\begin{array}{cc}
			1+\frac{b}{4\lambda}+\frac{b}{4}\Delta_m x_m & -\frac{b}{4}\nabla_m
			q_m \\
			\frac{b}{4}\nabla_m r_m & 1-\frac{b}{4\lambda}+\frac{b}{4}\Delta_m
			x_m \end{array}\right)\Psi_m. \label{sdSPtimeLP}
	\end{eqnarray}
	The consistency condition of linear system (\ref{sdSPtimeLP}), that
	is, $L_m \left(\frac{d}{dy}\Psi_m\right)=\frac{d}{dy}\left(L_m
	\Psi_m\right)$, yield semi-discrete SP equation (discrete in time)
	given by
	\begin{eqnarray}
		&&\left(\frac{1}{b}+\frac{1}{4}\Delta_m
		x_m\right)\frac{d}{dy}\Delta_m
		x_m+\frac{1}{8}\left(\frac{d}{dy}\nabla_m q_m\nabla_m r_m+\nabla_m
		q_m\frac{d}{dy}\nabla_m r_m\right)=0, \nonumber \\
		&&\left(\frac{1}{b}+\frac{1}{4}\Delta_m
		x_m\right)\frac{d}{dy}\Delta_m q_m-\frac{1}{4}\nabla_m
		q_m\frac{d}{dy}\nabla_m x_m=0, \nonumber \\
		&&\left(\frac{1}{b}+\frac{1}{4}\Delta_m
		x_m\right)\frac{d}{dy}\Delta_m r_m-\frac{1}{4}\nabla_m
		r_m\frac{d}{dy}\nabla_m x_m=0. \label{sdSPtime}
	\end{eqnarray}
	System of equation (\ref{sdSPtime}) and its associated linear system
	(\ref{sdSPtimeLP}) under the continuum limit will reduce to
	respective continuous counterparts (\ref{nsp07}) and (\ref{nsp08}),
	respectively. This time discretized version of SP equation is  a new
	addition in literature.
	
	\subsection{Space discretized version of SP equation}
	Similarly another semi-discrete version of SP equation can be
	obtained from (\ref{DSP1})-(\ref{DSP3}) under the continuum limit
	$b\rightarrow 0$ and $m\rightarrow \infty$ as
	\begin{eqnarray}
		\frac{d}{d\tau}\left( x_{n+1}-x_{n}\right) +\frac{1}{2}\left(
		q_{n+1}r_{n+1}-q_{n}r_{n}\right)  &=&0,  \nonumber \\
		\frac{d}{d\tau}\left( q_{n+1}-q_{n}\right) -\frac{1}{2}\left(
		x_{n+1}-x_{n}\right) \left( q_{n+1}+q_{n}\right)  &=&0,  \nonumber \\
		\frac{d}{d\tau}\left( r_{n+1}-r_{n}\right) -\frac{1}{2}\left(
		x_{n+1}-x_{n}\right) \left( r_{n+1}+r_{n}\right) &=&0.
		\label{GENsdSP}
	\end{eqnarray}
	Above system of equations arises as consistency condition of the
	following pair of difference-differential equations (also known as
	semi-discrete Lax pair)
	\begin{eqnarray}
		\Psi_{n+1}&=&\left(I_2+\lambda Q_n\right)\Psi_n, \nonumber\\
		\frac{d}{d
			\tau}\Psi_n&=&\left(\frac{1}{4\lambda}\sigma_3+\frac{1}{2}R_n\right)\Psi_n,\label{LPsd}
	\end{eqnarray}
	where $2 \times 2$ matrices $Q_n$ and $R_n$ are defined below
	\begin{eqnarray}
		Q_n=\left(\begin{array}{cc} \Delta_n x_n & \Delta_n q_n\\
			\Delta_n r_n & -\Delta_n x_n \end{array}\right),\quad \quad \quad
		\quad R_n=\left(\begin{array}{cc} 0 & -q_n\\
			r_n & 0 \end{array}\right). 
	\end{eqnarray}
	This was introduced by Feng and coworkers in their work
	\cite{feng2015integrable}. It can easily be verified that semi
	discrete SP equation (\ref{GENsdSP}) and its associated linear
	system (\ref{LPsd}) can be obtained from fully discrete SP equation
	(\ref{DSP1})-(\ref{DSP3}) and its discrete linear system (\ref{LP})
	under the limit $b\rightarrow 0$. Also under $n \rightarrow \infty$,
	above defined equation and linear system will reduce to their
	respective continuous counterparts. We have already investigated
	this system (\ref{GENsdSP}) under nonlocal symmetry reduction in our
	work \cite{hanif2020pt}.

\section{Discrete Darboux transformation}
Darboux transformation (DT) is a well-established mathematical
technique which has been used to compute multi-soliton solutions of
many integrable systems \cite{matveev1991darboux}. It is more or
less like a gauge transformation that relates two matrix-valued
solutions of the associated linear system via a Darboux matrix whose
application keeps the linear system and associated equations
covariant. The covariance of the system is then used to compute
higher-order nontrivial solutions under successive iterations of DT.
One-fold DT on the matrix-valued solution $\Psi_{n,m}$ of the
discrete Lax pair (\ref{LP}) is defined as:
\begin{equation}
	\Psi_{n,m}[1]=T^{[1]}_{n.m}\Psi_{n,m}, \label{DT1}
\end{equation}
where $T^{[1]}_{n.m}$ is the Darboux matrix and is given by
\begin{equation}
	T^{[1]}_{n.m}=\left( \begin{array}{cc}
		\lambda^{-1}+\alpha^{[0]}_{n,m} & \beta^{[0]}_{n,m}\\
		\gamma^{[0]}_{n,m} & \lambda^{-1}+\delta^{[0]}_{n,m}
	\end{array}\right),
\end{equation}
where $\alpha^{[0]}_{n,m}$, $\beta^{[0]}_{n,m}$,
$\gamma^{[0]}_{n,m}$ and $\delta^{[0]}_{n,m}$ can be determined by
implying following constraint
\begin{equation}
	T^{[1]}_{n,m}\Psi_{n,m}|_{\lambda\rightarrow
		\lambda_i}=\left|0\right>. \quad \quad \quad \quad \left(i=1,2
	\right) \label{T1}
\end{equation}
If we take $\Psi_{n,m}=\left(f_{n,m} \quad g_{n,m}\right)^T$ is a
column solution of difference-difference equation (\ref{LP}) at
$\lambda$, then for $i=1,2$, above expression (\ref{T1}) become
\begin{eqnarray*}
	&&\left( \begin{array}{cc}
		\lambda^{-1}_1+\alpha^{[0]}_{n,m} & \beta^{[0]}_{n,m}\\
		\gamma^{[0]}_{n,m} & \lambda^{-1}_1+\delta^{[0]}_{n,m}
	\end{array}\right)\left(\begin{array}{c}
		f^{(1)}_{n,m} \\
		g^{(1)}_{n,m} \end{array}\right)=\left(\begin{array}{c}
		0 \\
		0 \end{array}\right),\\
	&&\left( \begin{array}{cc}
		\lambda^{-1}_2+\alpha^{[0]}_{n,m} & \beta^{[0]}_{n,m}\\
		\gamma^{[0]}_{n,m} & \lambda^{-1}_2+\delta^{[0]}_{n,m}
	\end{array}\right)\left(\begin{array}{c}
		f^{(2)}_{n,m} \\
		g^{(2)}_{n,m} \end{array}\right)=\left(\begin{array}{c}
		0 \\
		0 \end{array}\right).
\end{eqnarray*}
We have to compute four unknown discrete-valued functions
$\alpha^{[0]}_{n,m}$, $\beta^{[0]}_{n,m}$, $\gamma^{[0]}_{n,m}$ and
$\delta^{[0]}_{n,m}$ from four linear equations. After rearranging
we have,
\begin{eqnarray*}
	&&\alpha^{[0]}_{n,m}f^{(1)}_{n,m}+\beta^{[0]}_{n,m}g^{(1)}_{n,m}=-\lambda^{-1}_1
	f^{(1)}_{n,m},\\
	&&\alpha^{[0]}_{n,m}f^{(2)}_{n,m}+\beta^{[0]}_{n,m}g^{(2)}_{n,m}=-\lambda^{-1}_2
	f^{(2)}_{n,m}, 
\end{eqnarray*}
and
\begin{eqnarray*}
	&&\gamma^{[0]}_{n,m}f^{(1)}_{n,m}+\delta^{[0]}_{n,m}g^{(1)}_{n,m}=-\lambda^{-1}_1
	g^{(1)}_{n,m}, \\
	&&\gamma^{[0]}_{n,m}f^{(2)}_{n,m}+\delta^{[0]}_{n,m}g^{(2)}_{n,m}=-\lambda^{-1}_2
	g^{(2)}_{n,m}. 
\end{eqnarray*}
By employing renowned technique of Cramer's rule we eventually
arrive at
\begin{eqnarray*}
	&& \alpha^{[0]}_{n,m}=-\frac{\left|\begin{array}{cc} \lambda^{-1}_1
			f^{(1)}_{n,m} & g^{(1)}_{n,m}\\
			\lambda^{-1}_2 f^{(2)}_{n,m} &
			g^{(2)}_{n,m}\end{array}\right|}{\left|\begin{array}{cc}
			f^{(1)}_{n,m} & g^{(1)}_{n,m}\\
			f^{(2)}_{n,m} & g^{(2)}_{n,m}\end{array}\right|}, \quad \quad
	\beta^{[0]}_{n,m}=-\frac{\left|\begin{array}{cc}
			f^{(1)}_{n,m} & \lambda^{-1}_1f^{(1)}_{n,m}\\
			f^{(2)}_{n,m} &
			\lambda^{-1}_2f^{(2)}_{n,m}\end{array}\right|}{\left|\begin{array}{cc}
			f^{(1)}_{n,m} & g^{(1)}_{n,m}\\
			f^{(2)}_{n,m} & g^{(2)}_{n,m}\end{array}\right|},  \\
	&& \gamma^{[0]}_{n,m}=-\frac{\left|\begin{array}{cc} \lambda^{-1}_1
			g^{(1)}_{n,m} & g^{(1)}_{n,m}\\
			\lambda^{-1}_2 g^{(2)}_{n,m} &
			g^{(2)}_{n,m}\end{array}\right|}{\left|\begin{array}{cc}
			f^{(1)}_{n,m} & g^{(1)}_{n,m}\\
			f^{(2)}_{n,m} & g^{(2)}_{n,m}\end{array}\right|}, \quad \quad
	\delta^{[0]}_{n,m}=-\frac{\left|\begin{array}{cc}
			f^{(1)}_{n,m} & \lambda^{-1}_1g^{(1)}_{n,m}\\
			f^{(2)}_{n,m} &
			\lambda^{-1}_2g^{(2)}_{n,m}\end{array}\right|}{\left|\begin{array}{cc}
			f^{(1)}_{n,m} & g^{(1)}_{n,m}\\
			f^{(2)}_{n,m} & g^{(2)}_{n,m}\end{array}\right|}. 
\end{eqnarray*}
Covariance of discrete linear system (\ref{LP}) under the action of
one-fold DT (\ref{DT1}), allows us to write
\begin{equation}
	\Psi_{n+1,m}[1]=\left(I_2+\lambda Q_{n,m}[1]\right)\Psi_{n,m}[1],
	\quad \quad
	\Psi_{n,m+1}[1]=\left(I_2+\frac{b}{4\lambda}\sigma_3+\frac{b}{4}
	R_{n,m}[1]\right)\Psi_{n,m}[1],
\end{equation}
that implies
\begin{eqnarray}
	&&x_{n,m}[1]=x_{n,m}+\frac{1}{2}\left(\alpha^{[0]}_{n,m}-\delta^{[0]}_{n,m}\right),
	\nonumber \\
	&&q_{n,m}[1]=q_{n,m}+\beta^{[0]}_{n,m}, \nonumber \\
	&&r_{n,m}[1]=r_{n,m}+\gamma^{[0]}_{n,m}. \label{1foldEXPGen}
\end{eqnarray}
The result (\ref{1foldEXPGen}) enables us to explicitly calculate
one-loop soliton solution of discrete SP equation
(\ref{DSP1a})-(\ref{DSP2b}). Similarly, the second iteration of DT
on the matrix-valued discrete function is defined as
\begin{equation}
	\Psi_{n,m}[2]\equiv T^{[2]}_{n.m}\Psi_{n,m}=\left( \begin{array}{cc}
		\lambda^{-2}+\lambda^{-1}\alpha^{[1]}_{n,m}+\alpha^{[0]}_{n,m} & \lambda^{-1}\beta^{[1]}_{n,m}+\beta^{[0]}_{n,m} \\
		\lambda^{-1}\gamma^{[1]}_{n,m}+\gamma^{[0]}_{n,m} &
		\lambda^{-2}+\lambda^{-1}\delta^{[1]}_{n,m}+\delta^{[0]}_{n,m}
	\end{array}\right)\Psi_{n,m}. \label{DT2}
\end{equation}
Covariance of discrete Lax pair (\ref{LP}) under DT (\ref{DT2})
demands
\begin{eqnarray}
	&&x_{n,m}[2]=x_{n,m}+\frac{1}{2}\left(\alpha^{[1]}_{n,m}-\delta^{[1]}_{n,m}\right),
	\nonumber \\
	&&q_{n,m}[2]=q_{n,m}+\beta^{[1]}_{n,m}, \nonumber \\
	&&r_{n,m}[2]=r_{n,m}+\gamma^{[1]}_{n,m}. \label{2foldEXPGen}
\end{eqnarray}
The unknown discrete valued functions $\alpha^{[k]}_{n,m}$,
$\beta^{[k]}_{n,m}$, $\gamma^{[k]}_{n,m}$ and $\delta^{[k]}_{n,m}$
($k=0,1$) can be calculated by using the constraint
$T^{[2]}_{n,m}\Psi_{n,m}|_{\lambda \rightarrow
	\lambda_i}=\left|0\right>$ where $i=1,2,3,4$. Under this condition,
we will arrive at following systems of equation expressed in matrix
form,
\begin{eqnarray*}
	\left(\begin{array}{cccc} f^{(1)}_{n,m} & g^{(1)}_{n,m} &
		\lambda^{-1}_1 f^{(1)}_{n,m} & \lambda^{-1}_1 g^{(1)}_{n,m} \\
		f^{(2)}_{n,m} & g^{(2)}_{n,m} & \lambda^{-1}_2 f^{(2)}_{n,m} &
		\lambda^{-1}_2 g^{(2)}_{n,m} \\
		f^{(3)}_{n,m} & g^{(3)}_{n,m} & \lambda^{-1}_3 f^{(3)}_{n,m} &
		\lambda^{-1}_3 g^{(3)}_{n,m} \\
		f^{(4)}_{n,m} & g^{(4)}_{n,m} & \lambda^{-1}_4 f^{(4)}_{n,m} &
		\lambda^{-1}_4 g^{(4)}_{n,m}
	\end{array}\right)
	\left(\begin{array}{c} \alpha^{[0]}_{n,m}\\
		\beta^{[0]}_{n,m}\\
		\alpha^{[1]}_{n,m}\\
		\beta^{[1]}_{n,m} \end{array}\right)&=&
	\left(\begin{array}{c} -\lambda_1^{-2}f^{(1)}_{n,m}\\
		-\lambda_2^{-2}f^{(2)}_{n,m}\\
		-\lambda_3^{-2}f^{(3)}_{n,m}\\
		-\lambda_4^{-2}f^{(4)}_{n,m} \end{array}\right), \\
	\left(\begin{array}{cccc} f^{(1)}_{n,m} & g^{(1)}_{n,m} &
		\lambda^{-1}_1 f^{(1)}_{n,m} & \lambda^{-1}_1 g^{(1)}_{n,m} \\
		f^{(2)}_{n,m} & g^{(2)}_{n,m} & \lambda^{-1}_2 f^{(2)}_{n,m} &
		\lambda^{-1}_2 g^{(2)}_{n,m} \\
		f^{(3)}_{n,m} & g^{(3)}_{n,m} & \lambda^{-1}_3 f^{(3)}_{n,m} &
		\lambda^{-1}_3 g^{(3)}_{n,m} \\
		f^{(4)}_{n,m} & g^{(4)}_{n,m} & \lambda^{-1}_4 f^{(4)}_{n,m} &
		\lambda^{-1}_4 g^{(4)}_{n,m}
	\end{array}\right)
	\left(\begin{array}{c} \gamma^{[0]}_{n,m}\\
		\delta^{[0]}_{n,m}\\
		\gamma^{[1]}_{n,m}\\
		\delta^{[1]}_{n,m} \end{array}\right)&=&
	\left(\begin{array}{c} -\lambda_1^{-2}g^{(1)}_{n,m}\\
		-\lambda_2^{-2}g^{(2)}_{n,m}\\
		-\lambda_3^{-2}g^{(3)}_{n,m}\\
		-\lambda_4^{-2}g^{(4)}_{n,m} \end{array}\right).
\end{eqnarray*}
Above systems can be easily solved for unknown discrete functions by
using a trivial technique of Cramer's method. Eventually, we obtain
\begin{eqnarray*}
	\alpha^{[1]}_{n,m}&=& -\frac{\mbox{det}\left(\begin{array}{cccc}
			f^{(1)}_{n,m} & g^{(1)}_{n,m} &
			\lambda^{-2}_1 f^{(1)}_{n,m} & \lambda^{-1}_1 g^{(1)}_{n,m} \\
			f^{(2)}_{n,m} & g^{(2)}_{n,m} & \lambda^{-2}_2 f^{(2)}_{n,m} &
			\lambda^{-1}_2 g^{(2)}_{n,m} \\
			f^{(3)}_{n,m} & g^{(3)}_{n,m} & \lambda^{-2}_3 f^{(3)}_{n,m} &
			\lambda^{-1}_3 g^{(3)}_{n,m} \\
			f^{(4)}_{n,m} & g^{(4)}_{n,m} & \lambda^{-2}_4 f^{(4)}_{n,m} &
			\lambda^{-1}_4 g^{(4)}_{n,m}
		\end{array}\right)}
	{\mbox{det}\left(\begin{array}{cccc} f^{(1)}_{n,m} & g^{(1)}_{n,m} &
			\lambda^{-1}_1 f^{(1)}_{n,m} & \lambda^{-1}_1 g^{(1)}_{n,m} \\
			f^{(2)}_{n,m} & g^{(2)}_{n,m} & \lambda^{-1}_2 f^{(2)}_{n,m} &
			\lambda^{-1}_2 g^{(2)}_{n,m} \\
			f^{(3)}_{n,m} & g^{(3)}_{n,m} & \lambda^{-1}_3 f^{(3)}_{n,m} &
			\lambda^{-1}_3 g^{(3)}_{n,m} \\
			f^{(4)}_{n,m} & g^{(4)}_{n,m} & \lambda^{-1}_4 f^{(4)}_{n,m} &
			\lambda^{-1}_4 g^{(4)}_{n,m}
		\end{array}\right)}, \\
	\beta^{[1]}_{n,m}&=& -\frac{\mbox{det}\left(\begin{array}{cccc}
			f^{(1)}_{n,m} & g^{(1)}_{n,m} &
			\lambda^{-1}_1 f^{(1)}_{n,m} & \lambda^{-2}_1 f^{(1)}_{n,m} \\
			f^{(2)}_{n,m} & g^{(2)}_{n,m} & \lambda^{-1}_2 f^{(2)}_{n,m} &
			\lambda^{-2}_2 f^{(2)}_{n,m} \\
			f^{(3)}_{n,m} & g^{(3)}_{n,m} & \lambda^{-1}_3 f^{(3)}_{n,m} &
			\lambda^{-2}_3 f^{(3)}_{n,m} \\
			f^{(4)}_{n,m} & g^{(4)}_{n,m} & \lambda^{-1}_4 f^{(4)}_{n,m} &
			\lambda^{-2}_4 f^{(4)}_{n,m}
		\end{array}\right)}
	{\mbox{det}\left(\begin{array}{cccc} f^{(1)}_{n,m} & g^{(1)}_{n,m} &
			\lambda^{-1}_1 f^{(1)}_{n,m} & \lambda^{-1}_1 g^{(1)}_{n,m} \\
			f^{(2)}_{n,m} & g^{(2)}_{n,m} & \lambda^{-1}_2 f^{(2)}_{n,m} &
			\lambda^{-1}_2 g^{(2)}_{n,m} \\
			f^{(3)}_{n,m} & g^{(3)}_{n,m} & \lambda^{-1}_3 f^{(3)}_{n,m} &
			\lambda^{-1}_3 g^{(3)}_{n,m} \\
			f^{(4)}_{n,m} & g^{(4)}_{n,m} & \lambda^{-1}_4 f^{(4)}_{n,m} &
			\lambda^{-1}_4 g^{(4)}_{n,m}
		\end{array}\right)}, \\
	\gamma^{[1]}_{n,m}&=& -\frac{\mbox{det}\left(\begin{array}{cccc}
			f^{(1)}_{n,m} & g^{(1)}_{n,m} &
			\lambda^{-2}_1 g^{(1)}_{n,m} & \lambda^{-1}_1 g^{(1)}_{n,m} \\
			f^{(2)}_{n,m} & g^{(2)}_{n,m} & \lambda^{-2}_2 g^{(2)}_{n,m} &
			\lambda^{-1}_2 g^{(2)}_{n,m} \\
			f^{(3)}_{n,m} & g^{(3)}_{n,m} & \lambda^{-2}_3 g^{(3)}_{n,m} &
			\lambda^{-1}_3 g^{(3)}_{n,m} \\
			f^{(4)}_{n,m} & g^{(4)}_{n,m} & \lambda^{-2}_4 g^{(4)}_{n,m} &
			\lambda^{-1}_4 g^{(4)}_{n,m}
		\end{array}\right)}
	{\mbox{det}\left(\begin{array}{cccc} f^{(1)}_{n,m} & g^{(1)}_{n,m} &
			\lambda^{-1}_1 f^{(1)}_{n,m} & \lambda^{-1}_1 g^{(1)}_{n,m} \\
			f^{(2)}_{n,m} & g^{(2)}_{n,m} & \lambda^{-1}_2 f^{(2)}_{n,m} &
			\lambda^{-1}_2 g^{(2)}_{n,m} \\
			f^{(3)}_{n,m} & g^{(3)}_{n,m} & \lambda^{-1}_3 f^{(3)}_{n,m} &
			\lambda^{-1}_3 g^{(3)}_{n,m} \\
			f^{(4)}_{n,m} & g^{(4)}_{n,m} & \lambda^{-1}_4 f^{(4)}_{n,m} &
			\lambda^{-1}_4 g^{(4)}_{n,m}
		\end{array}\right)}, \\
	\delta^{[1]}_{n,m}&=& -\frac{\mbox{det}\left(\begin{array}{cccc}
			f^{(1)}_{n,m} & g^{(1)}_{n,m} &
			\lambda^{-1}_1 f^{(1)}_{n,m} & \lambda^{-2}_1 g^{(1)}_{n,m} \\
			f^{(2)}_{n,m} & g^{(2)}_{n,m} & \lambda^{-1}_2 f^{(2)}_{n,m} &
			\lambda^{-2}_2 g^{(2)}_{n,m} \\
			f^{(3)}_{n,m} & g^{(3)}_{n,m} & \lambda^{-1}_3 f^{(3)}_{n,m} &
			\lambda^{-2}_3 g^{(3)}_{n,m} \\
			f^{(4)}_{n,m} & g^{(4)}_{n,m} & \lambda^{-1}_4 f^{(4)}_{n,m} &
			\lambda^{-2}_4 g^{(4)}_{n,m}
		\end{array}\right)}
	{\mbox{det}\left(\begin{array}{cccc} f^{(1)}_{n,m} & g^{(1)}_{n,m} &
			\lambda^{-1}_1 f^{(1)}_{n,m} & \lambda^{-1}_1 g^{(1)}_{n,m} \\
			f^{(2)}_{n,m} & g^{(2)}_{n,m} & \lambda^{-1}_2 f^{(2)}_{n,m} &
			\lambda^{-1}_2 g^{(2)}_{n,m} \\
			f^{(3)}_{n,m} & g^{(3)}_{n,m} & \lambda^{-1}_3 f^{(3)}_{n,m} &
			\lambda^{-1}_3 g^{(3)}_{n,m} \\
			f^{(4)}_{n,m} & g^{(4)}_{n,m} & \lambda^{-1}_4 f^{(4)}_{n,m} &
			\lambda^{-1}_4 g^{(4)}_{n,m}
		\end{array}\right)}. 
\end{eqnarray*}
Again, the expression (\ref{2foldEXPGen}) gives the interactions of
loop-loop, loop-antiloop soliton solutions and breather solution of
discrete SP equation (\ref{DSP1a})-(\ref{DSP2b}).

The $N$-fold DT can be generalized as
\begin{equation}
	\Psi_{n,m} \lbrack N]=T^{[N]}_{n,m}\Psi_{n,m} ,  \label{nsp10}
\end{equation}%
with
\begin{equation}
	T^{[N]}_{n,m}=\left(
	\begin{array}{cc}
		\lambda ^{-N}+\sum\limits_{k=0}^{N-1}\alpha^{[k]}_{n,m}\lambda ^{-k} & \sum\limits_{k=0}^{N-1}\beta_{n,m}^{[k]}%
		\lambda ^{-k} \\
		\sum\limits_{k=0}^{N-1}\gamma_{n,m}^{[k]}\lambda ^{-k} & \lambda
		^{-N}+\sum\limits_{k=0}^{N-1}\delta_{n,m}^{[k]}\lambda ^{-k}%
	\end{array}%
	\right).  \label{nsp11}
\end{equation}
Under the action of DT (\ref{nsp10}) the discrete dynamical variables transform as%
\begin{eqnarray}
	x_{n,m}[N] &=&x_{n,m}+\frac{1}{2}\left(\alpha_{n,m}^{[N-1]}-\delta_{n,m}^{[N-1]}\right),  \nonumber \\
	q_{n,m}[N] &=&q_{n,m}+\beta_{n,m}^{[N-1]},  \nonumber \\
	r_{n,m}[N] &=&r_{n,m}+\gamma_{n,m}^{[N-1]}.  \label{nsp12}
\end{eqnarray}
The unknowns $\alpha_{n,m}^{[k]}$, $\beta_{n,m}^{[k]}$,
$\gamma_{n,m}^{[k]}$ and $\delta_{n,m}^{[k]}$ $\left( 0\leq k\leq
N-1\right) $ can be uniquely determined by requiring%
\begin{equation}
	T_{n,m}^{[N]}\Psi_{n,m}|_{\lambda \rightarrow \lambda
		_{i}}=\left|0\right>, \quad \quad \quad \quad \quad (i=1,2,\cdots ,
	2N) \label{nsp13}
\end{equation}
which yields system of equations%
\begin{eqnarray}
	\sum\limits_{k=0}^{N-1}\alpha_{n,m}^{[k]} \lambda _{i}^{-k} %
	f_{n,m}^{(i)}+\sum\limits_{k=0}^{N-1}\beta_{n,m}^{[k]} \lambda
	_{i}^{-k} g_{n,m}^{(i)}
	&=&-\lambda _{i}^{-N}f_{n,m}^{(i)},  \nonumber \\
	\sum\limits_{k=0}^{N-1}\gamma_{n,m}^{[k]} \lambda _{i}^{-k} %
	f_{n,m}^{(i)}+\sum\limits_{k=0}^{N-1}\delta_{n,m}^{[k]} \lambda
	_{i}^{-k} g_{n,m}^{(i)} &=&-\lambda _{i}^{-N}g_{n,m}^{(i)},
	\label{nsp14}
\end{eqnarray}
where $\lambda _{i}$ $\left( 1\leq i\leq 2N\right) $ are spectral
parameters and $\Psi _{n,m}^{(i)}\left( \lambda _{i}\right) =\left(
f_{n,m}^{(i)} \quad g_{n,m}^{(i)}\right) ^{T}$ are the vector
solutions of system (\ref{LP}). The linear system (\ref{nsp14}) can
be solved by using Cramer's rule, the coefficient matrices provide
the
following results%
\begin{eqnarray}
	\alpha_{n,m}^{[N-1]} &=&\frac{\Delta _{\alpha_{n,m}}^{[N-1]}}{\Delta ^{[N-1]}},\quad \quad \beta_{n,m}^{[N-1]}=\frac{%
		\Delta _{\beta_{n,m}}^{[N-1]}}{\Delta ^{[N-1]}},  \nonumber \\
	\gamma_{n,m}^{[N-1]} &=&\frac{\Delta _{\gamma_{n,m}}^{[N-1]}}{\Delta ^{[N-1]}},\quad \quad \delta_{n,m}^{[N-1]}=\frac{%
		\Delta _{\delta_{n,m}}^{[N-1]}}{\Delta ^{[N-1]}}, \label{nsp15}
\end{eqnarray}
with%
\begin{eqnarray*}
	\Delta^{[N-1]}&=&\mbox{det}\left(\begin{array}{ccccccc}
		f^{(i)}_{n,m} & g^{(i)}_{n,m} & \cdots & \lambda^{-N+2}_i
		f^{(i)}_{n,m} & \lambda^{-N+2}_i g^{(i)}_{n,m} & \lambda^{-N+1}_i
		f^{(i)}_{n,m} & \lambda^{-N+1}_i g^{(i)}_{n,m}
	\end{array}\right),  \\
	\Delta_{\alpha_{n,m}}^{[N-1]}&=&\mbox{det}\left(\begin{array}{ccccccc}
		f^{(i)}_{n,m} & g^{(i)}_{n,m} & \cdots & \lambda^{-N+2}_i
		f^{(i)}_{n,m} & \lambda^{-N+2}_i g^{(i)}_{n,m} & \lambda^{-N}_i
		f^{(i)}_{n,m} & \lambda^{-N+1}_i g^{(i)}_{n,m}
	\end{array}\right), \\
	\Delta_{\beta_{n,m}}^{[N-1]}&=&\mbox{det}\left(\begin{array}{ccccccc}
		f^{(i)}_{n,m} & g^{(i)}_{n,m} & \cdots & \lambda^{-N+2}_i
		f^{(i)}_{n,m} & \lambda^{-N+2}_i g^{(i)}_{n,m} & \lambda^{-N+1}_i
		f^{(i)}_{n,m} & \lambda^{-N}_i f^{(i)}_{n,m}
	\end{array}\right),  \\
	\Delta_{\gamma_{n,m}}^{[N-1]}&=&\mbox{det}\left(\begin{array}{ccccccc}
		f^{(i)}_{n,m} & g^{(i)}_{n,m} & \cdots & \lambda^{-N+2}_i
		f^{(i)}_{n,m} & \lambda^{-N+2}_i g^{(i)}_{n,m} & \lambda^{-N}_i
		g^{(i)}_{n,m} & \lambda^{-N+1}_i g^{(i)}_{n,m}
	\end{array}\right), \\
	\Delta_{\delta_{n,m}}^{[N-1]}&=&\mbox{det}\left(\begin{array}{ccccccc}
		f^{(i)}_{n,m} & g^{(i)}_{n,m} & \cdots & \lambda^{-N+2}_i
		f^{(i)}_{n,m} & \lambda^{-N+2}_i g^{(i)}_{n,m} & \lambda^{-N+1}_i
		f^{(i)}_{n,m} & \lambda^{-N}_i g^{(i)}_{n,m}
	\end{array}\right).
\end{eqnarray*}
The higher-order loop-soliton solutions of discrete SP equation
(\ref{DSP1a})-(\ref{DSP2b}) can be computed throught the results
(\ref{nsp12}) and (\ref{nsp15}). In what follows next, we will
compute first two nontrivial solution of the fully discrete SP
equation (\ref{DSP1a})-(\ref{DSP2b}). We also apply continuum limits
to relate our results with already known solutions of semi-discrete
and continuous SP equations.

\section{Explicit solutions}
In this section explicit expressions of first two nontrivial
solutions of discrete SP equation (\ref{DSP1a})-(\ref{DSP2b}) will
be computed. Consider a seed solution $x_{n,m}=na$ and
$q_{n,m}=r_{n,m}=0$, the associated Lax pair (\ref{LP}) will become
\begin{equation}
	\Psi_{n+1,m}=\left(\begin{array}{cc} 1+a\lambda & 0 \\
		0 & 1-a\lambda \end{array}\right)\Psi_{n,m}, \quad \quad \quad \Psi_{n,m+1}=\left(\begin{array}{cc} 1+\frac{b}{4\lambda} & 0 \\
		0 & 1-\frac{b}{4\lambda} \end{array}\right)\Psi_{n,m}, \label{abc}
\end{equation}
where $\Psi_{n,m}=\left(f_{n,m} \quad g_{n,m}\right)^T$. The
solution of linear system of difference-difference equations
(\ref{abc}) becomes
\begin{equation}
	f_{n,m}=A\left(1+a\lambda\right)^n
	\left(1+\frac{b}{4\lambda}\right)^m, \quad \quad \quad
	g_{n,m}=B\left(1-a\lambda\right)^n
	\left(1-\frac{b}{4\lambda}\right)^m, \label{XYnm1}
\end{equation}
where $A$ and $B$ are the constants. The particular column solutions
at $\lambda=\lambda_k$ are defined as
\begin{equation}
	f^{(i)}_{n,m}=A_i\left(1+a\lambda_i\right)^n
	\left(1+\frac{b}{4\lambda_i}\right)^m, \quad \quad \quad
	g^{(i)}_{n,m}=B_i\left(1-a\lambda_i\right)^n
	\left(1-\frac{b}{4\lambda_i}\right)^m. \label{XYnm}
\end{equation}
The reduction
requirement $q_{n,m}=-r_{n,m}$ will be realized when
\begin{equation}
	\lambda_{2l}=-\lambda_{2l-1}, \quad \quad \quad f^{(2l)}_{n,m}=\pm
	g^{(2l-1)}_{n,m}, \quad \quad \quad g^{(2l)}_{n,m}=\mp
	f^{(2l-1)}_{n,m}. \label{RedReq1}
\end{equation}

\subsection{First-order nontrivial loop soliton solution}
\begin{figure}
	\centering
	\includegraphics[width=80mm]{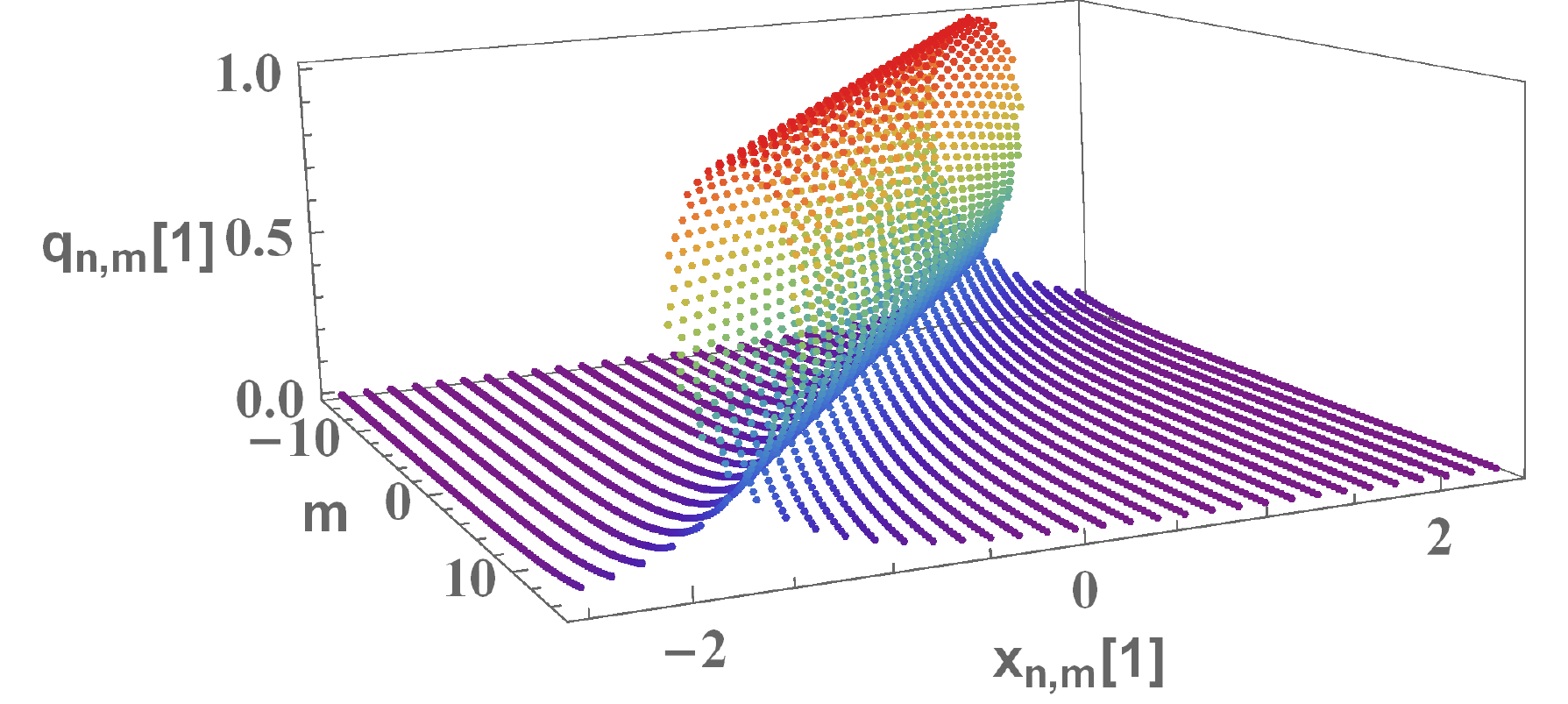}
	\caption{Loop soliton solution for fully discrete SP equation (\ref{DSP1a})-(\ref{DSP2b}).}
	\label{Local1}
\end{figure}

The expression (\ref{1foldEXPGen}) gives us explicit expression for
one-loop soliton solutions as
\begin{eqnarray}
	&&x_{n,m}[1]=an-\frac{1}{2}\left(\frac{\left|\begin{array}{cc}
			\lambda^{-1}_1
			f^{(1)}_{n,m} & g^{(1)}_{n,m}\\
			\lambda^{-1}_2 f^{(2)}_{n,m} &
			g^{(2)}_{n,m}\end{array}\right|}{\left|\begin{array}{cc}
			f^{(1)}_{n,m} & g^{(1)}_{n,m}\\
			f^{(2)}_{n,m} & g^{(2)}_{n,m}\end{array}\right|}-
	\frac{\left|\begin{array}{cc} f^{(1)}_{n,m} & \lambda^{-1}_1
			g^{(1)}_{n,m}\\
			f^{(2)}_{n,m} & \lambda^{-1}_2
			g^{(2)}_{n,m}\end{array}\right|}{\left|\begin{array}{cc}
			f^{(1)}_{n,m} & g^{(1)}_{n,m}\\
			f^{(2)}_{n,m} & g^{(2)}_{n,m}\end{array}\right|}\right), \nonumber
	\\
	&&q_{n,m}[1]=-\frac{\left|\begin{array}{cc} f^{(1)}_{n,m} &
			\lambda^{-1}_1
			f^{(1)}_{n,m}\\
			f^{(2)}_{n,m} & \lambda^{-1}_2
			f^{(2)}_{n,m}\end{array}\right|}{\left|\begin{array}{cc}
			f^{(1)}_{n,m} & g^{(1)}_{n,m}\\
			f^{(2)}_{n,m} & g^{(2)}_{n,m}\end{array}\right|}.\label{1FEXP}
\end{eqnarray}
Substituting $f^{(i)}_{n,m}$ and $g^{(i)}_{n,m}$ from (\ref{XYnm1})
and also applying reduction constraints (\ref{RedReq1}) in above
expressions we get,
\scriptsize
\begin{eqnarray}
	x_{n,m}\left[ 1\right]  &=&an-\frac{1}{\lambda _{1}}\left( 1-\frac{%
		2A_{2}B_{1}}{A_{2}B_{1}-A_{1}B_{2}\left( 4\lambda _{1}-b\right)
		^{-2m}\left( 4\lambda _{1}+b\right) ^{2m}\left( 1-a\lambda
		_{1}\right) ^{-2n}\left(
		1+a\lambda _{1}\right) ^{2n}}\right),   \nonumber \\
	q_{n,m}\left[ 1\right]  &=&\frac{2A_{1}A_{2}\left(
		1-\frac{b^{2}}{16\lambda _{1}^{2}}\right) ^{m}\left( 1-a^{2}\lambda
		_{1}^{2}\right) ^{n}}{\lambda _{1}\left( A_{1}B_{2}\left(
		1+\frac{b}{4\lambda _{1}}\right) ^{2m}\left(
		1+a\lambda _{1}\right) ^{2n}-A_{2}B_{1}\left( 1-\frac{b}{4\lambda _{1}}%
		\right) ^{2m}\left( 1-a\lambda _{1}\right) ^{2n}\right) }.
	\label{OneFold1}
\end{eqnarray}
\normalsize
Equation (\ref{OneFold1}) corresponds to one-soliton solution and
the parametric correlation leads us to the loop solutions of fully
discrete SP equation (\ref{DSP1a})-(\ref{DSP2b}) for
$A_{2l}=A_{2l-1}=1$ and $B_{2l-1}=-B_{2l}=1$. Profile of loop
soliton solution for $\lambda _{1}=-1$ and $a=b=0.5$ is shown in
Fig. (\ref{Local1}). Under the continuum limit as $a\rightarrow 0,$
$n\rightarrow \infty $
and $y=na$, expression (\ref{OneFold1}) gives the solution of time-discrete SP equation (\ref{sdSPtime})%
\begin{eqnarray}
	x_{m}\left( y\right) \left[ 1\right]  &=&y-\frac{1}{\lambda _{1}}\left( 1-%
	\frac{2}{1+e^{4\lambda _{1}y}\left( 4\lambda _{1}-b\right)
		^{-2m}\left(
		4\lambda _{1}+b\right) ^{2m}}\right),   \nonumber \\
	q_{m}\left( y\right) \left[ 1\right]  &=&-\frac{2e^{2\lambda _{1}y}\left( 1-%
		\frac{b^{2}}{16\lambda _{1}^{2}}\right) ^{m}}{\lambda _{1}\left( \left( 1-%
		\frac{b}{4\lambda _{1}}\right) ^{2m}+e^{4\lambda _{1}y}\left( 1+\frac{b}{%
			4\lambda _{1}}\right) ^{2m}\right) }.  \label{OneFold2}
\end{eqnarray}
Likewise, the continuum limit as $b\rightarrow 0,$ $m\rightarrow
\infty $ and $\tau =mb$, expression (\ref{OneFold1}) yields the
solution of space-discrete
SP equation \cite{feng2015integrable, hanif2020pt}%
\begin{eqnarray}
	x_{n}\left( \tau \right) \left[ 1\right]  &=&an-\frac{1}{\lambda
		_{1}}\left( 1-\frac{2}{1+e^{\frac{\tau }{\lambda _{1}}}\left(
		1-a\lambda _{1}\right)
		^{-2n}\left( 1+a\lambda _{1}\right) ^{2n}}\right),   \nonumber \\
	q_{n}\left( \tau \right) \left[ 1\right]  &=&-\frac{2e^{\frac{\tau }{%
				2\lambda _{1}}}\left( 1-a^{2}\lambda _{1}^{2}\right) ^{n}}{\lambda
		_{1}\left( \left( 1-a\lambda _{1}\right) ^{2n}+e^{\frac{\tau }{\lambda _{1}}%
		}\left( 1+a\lambda _{1}\right) ^{2n}\right) },  \label{OneFold3}
\end{eqnarray}
while, again under the limit as $a\rightarrow 0,$ $n\rightarrow
\infty $ and $y=na$ , the above expression (\ref{OneFold3}) reduces to%
\begin{eqnarray}
	x\left( y,\tau \right) \left[ 1\right]  &=&x-\frac{1}{\lambda
		_{1}}\tanh
	\left( 2y\lambda _{1}+\frac{\tau }{2\lambda _{1}}\right),   \nonumber \\
	q\left( y,t\right) \left[ 1\right]  &=&-\frac{1}{\lambda _{1}}\sec
	\left( 2y\lambda _{1}+\frac{\tau }{2\lambda _{1}}\right),
	\label{OneFold4}
\end{eqnarray}
which represents the loop soliton solution for the continuous SP
equation (\ref{nsp07a}). If we apply the continuum limit $b\rightarrow
0,$ $m\rightarrow \infty $ and $\tau =mb$, expression
(\ref{OneFold2}) will reduce to (\ref{OneFold4}).

\subsection{Second-order nontrivial solutions}
\begin{figure*}
	\centering
	\includegraphics[width=80mm]{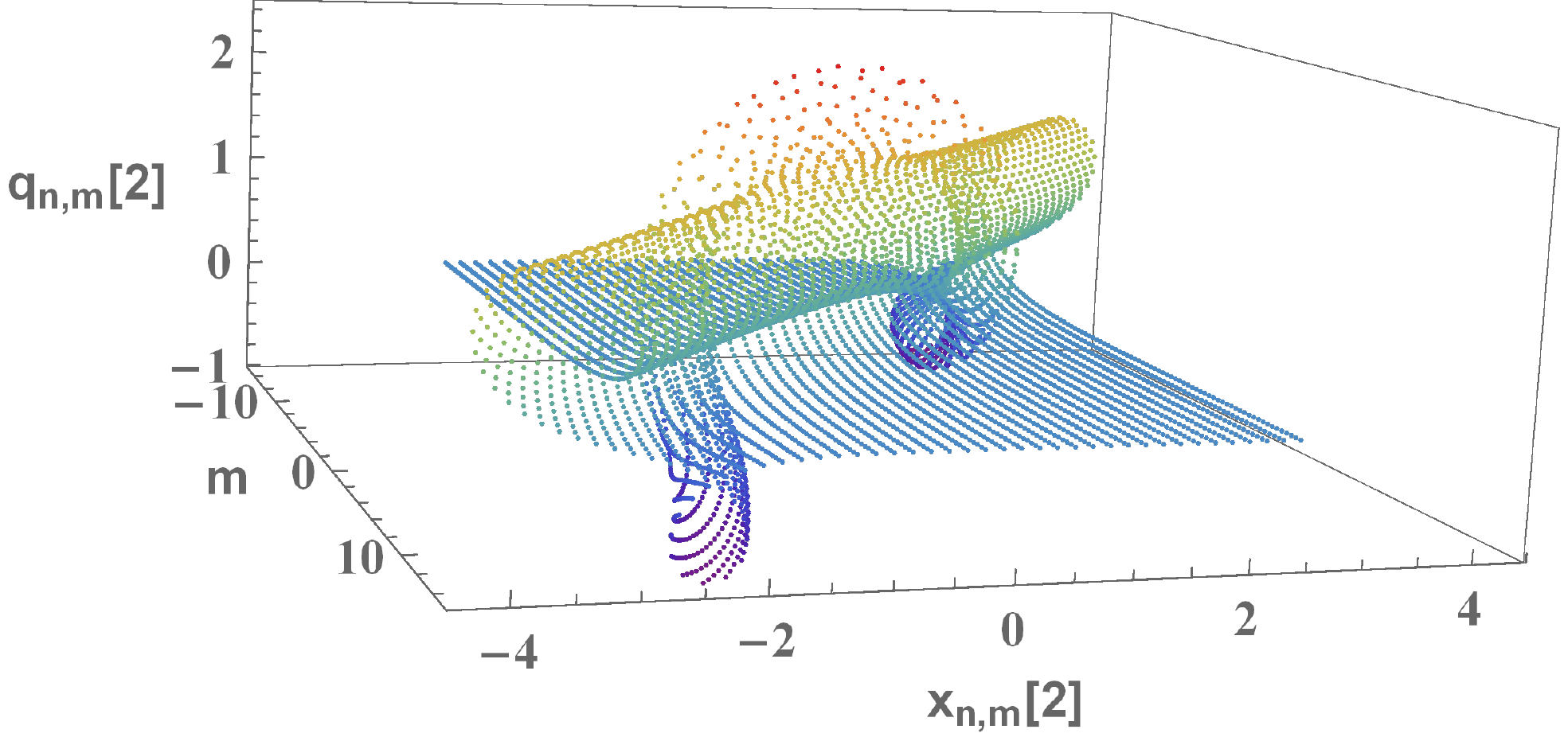}
	\caption{Loop and anti-loop interactions for fully discrete SP equation (\ref{DSP1a})-(\ref{DSP2b}).}
	\label{Local2Fa}
\end{figure*}

\begin{figure}
	\centering
	\includegraphics[width=80mm]{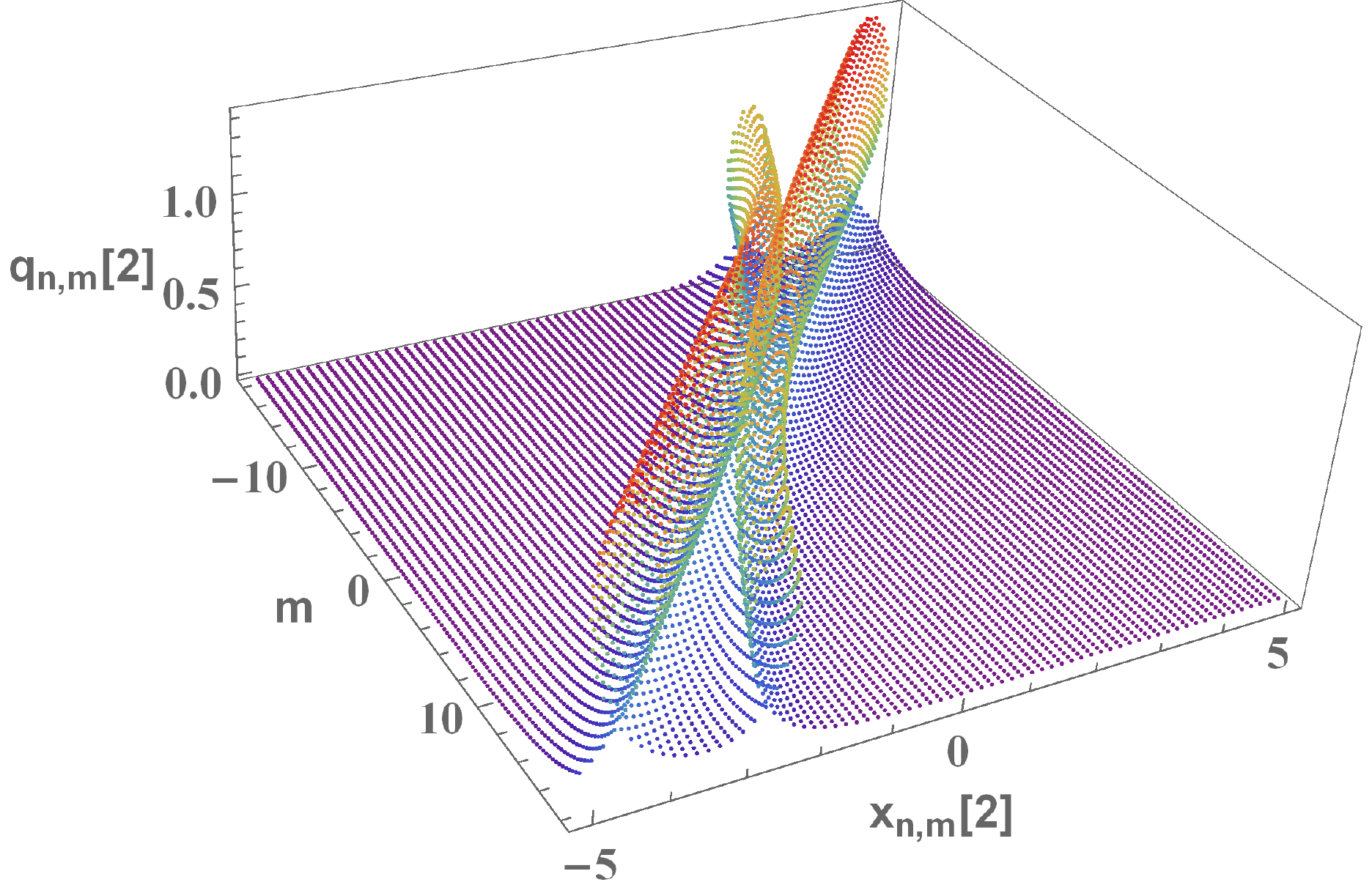}
	\caption{Loop-loop interactions for fully discrete SP equation (\ref{DSP1a})-(\ref{DSP2b}).}
	\label{Local2Fb}
\end{figure}

Equation (\ref{2foldEXPGen}) gives the explicit form of two-loop
soliton solutions for $A_{2l}=A_{2l-1}=1$ and $B_{2l-1}=-B_{2l}=1$,
$\lambda_2=-\lambda_1$ and $\lambda_4=-\lambda_3$ as
\begin{eqnarray}
	&&x_{n,m}[2]=\frac{(\lambda^2_1-\lambda^3_3) \left(-\lambda_1
		\Gamma^{(\alpha_1)}_{n,m} \Gamma^{(\beta_2)}_{n,m}+
		\lambda_3 \Gamma^{(\alpha_2)}_{n,m}\Gamma^{(\beta_1)}_{n,m}
		\right)}
	{\lambda_1
		\lambda_3 \left(-8\lambda_1\lambda_3\Gamma^{(a)}_{n,m}+\left(\lambda_1^2-\lambda_3^2\right)\Gamma^{(s)}_{n,m}
		+\left(\lambda_1^2+\lambda_3^2\right)\Gamma^{(d)}_{n,m}\right)},\nonumber \\
	&&q_{n,m}[2]=-\frac{2 (\lambda^2_1-\lambda^2_3) \left(\lambda_1
		\Gamma^{(\alpha_1)}_{n,m}\Gamma^{(\gamma_2)}_{n,m}-\lambda_3\Gamma^{(\beta_1)}_{n,m}\Gamma^{(\gamma_1)}_{n,m}
		\right)}{\lambda_1
		\lambda_3 \left(-8\lambda_1\lambda_3\Gamma^{(a)}_{n,m}+\left(\lambda_1^2-\lambda_3^2\right)\Gamma^{(s)}_{n,m}
		+\left(\lambda_1^2+\lambda_3^2\right)\Gamma^{(d)}_{n,m}\right)},
	\label{2foldEXP}
\end{eqnarray}
with
\begin{eqnarray*}
	\Gamma^{(\alpha_1)}_{n,m}&=&\left(1+a\lambda_1\right)^{2n}
	\left(1+\frac{b}{4\lambda_1}\right)^{2m}+
	\left(1-a\lambda_1\right)^{2n}
	\left(1-\frac{b}{4\lambda_1}\right)^{2m}, \\
	\Gamma^{(\alpha_2)}_{n,m}&=&\left(1+a\lambda_1\right)^{2n}
	\left(1+\frac{b}{4\lambda_1}\right)^{2m}-
	\left(1-a\lambda_1\right)^{2n}
	\left(1-\frac{b}{4\lambda_1}\right)^{2m}, \\
	\Gamma^{(\beta_1)}_{n,m}&=&\left(1+a\lambda_3\right)^{2n}
	\left(1+\frac{b}{4\lambda_3}\right)^{2m}+
	\left(1-a\lambda_3\right)^{2n}
	\left(1-\frac{b}{4\lambda_3}\right)^{2m}, \\
	\Gamma^{(\beta_2)}_{n,m}&=&\left(1+a\lambda_3\right)^{2n}
	\left(1+\frac{b}{4\lambda_3}\right)^{2m}-
	\left(1-a\lambda_3\right)^{2n}
	\left(1-\frac{b}{4\lambda_3}\right)^{2m}, \\
	\Gamma^{(a)}_{n,m}&=&\left(1+a\lambda_1\right)^{n}
	\left(1+\frac{b}{4\lambda_1}\right)^{m}\left(1+a\lambda_3\right)^{n}
	\left(1+\frac{b}{4\lambda_3}\right)^{m}
	\\
	&&\left(1-a\lambda_1\right)^{n}
	\left(1-\frac{b}{4\lambda_1}\right)^{m}\left(1-a\lambda_3\right)^{n}
	\left(1-\frac{b}{4\lambda_3}\right)^{m},\\
	\Gamma^{(s)}_{n,m}&=&\left(1+a\lambda_1\right)^{2n}
	\left(1+\frac{b}{4\lambda_1}\right)^{2m}\left(1+a\lambda_3\right)^{2n}
	\left(1+\frac{b}{4\lambda_3}\right)^{2m}+
	\\
	&&\left(1-a\lambda_1\right)^{2n}
	\left(1-\frac{b}{4\lambda_1}\right)^{2m}\left(1-a\lambda_3\right)^{2n}
	\left(1-\frac{b}{4\lambda_3}\right)^{2m},\\
	\Gamma^{(d)}_{n,m}&=&\left(1+a\lambda_1\right)^{2n}
	\left(1+\frac{b}{4\lambda_1}\right)^{2m}\left(1-a\lambda_3\right)^{2n}
	\left(1-\frac{b}{4\lambda_3}\right)^{2m}+
	\\
	&&\left(1+a\lambda_3\right)^{2n}
	\left(1+\frac{b}{4\lambda_3}\right)^{2m}\left(1-a\lambda_1\right)^{2n}
	\left(1-\frac{b}{4\lambda_1}\right)^{2m},\\
	\Gamma^{(\gamma_2)}_{n,m}&=&\left(1+a\lambda_3\right)^{n}
	\left(1+\frac{b}{4\lambda_3}\right)^{m}\left(1-a\lambda_3\right)^{n}
	\left(1-\frac{b}{4\lambda_3}\right)^{m},\\
	\Gamma^{(\gamma_1)}_{n,m}&=&\left(1+a\lambda_1\right)^{n}
	\left(1+\frac{b}{4\lambda_1}\right)^{m}\left(1-a\lambda_1\right)^{n}
	\left(1-\frac{b}{4\lambda_1}\right)^{m}.
\end{eqnarray*}
There exist two types of loop soliton interactions. The first one is
a loop-antiloop interface that is an attractive process in which a
loop and an antiloop while crossing form a spiral formation and
split in their respective course as shown in the Fig.
(\ref{Local2Fa}). This profile is obtained for $\lambda _{1}=-1$,
$\lambda _{3}=-0.7$ and $a=b=0.5$. The second one is a loop-loop
interface that is a repulsive progression in such collisions as two
loops approach they repel each other and also exchange their
energies and diverged as shown in the Fig. (\ref{Local2Fb}). This
profile is obtained for $\lambda _{1}=1$, $\lambda _{3}=-0.7$ and
$a=b=0.5$.

\subsection{Breather solutions}
\begin{figure*}
	\centering
	\includegraphics[width=70mm]{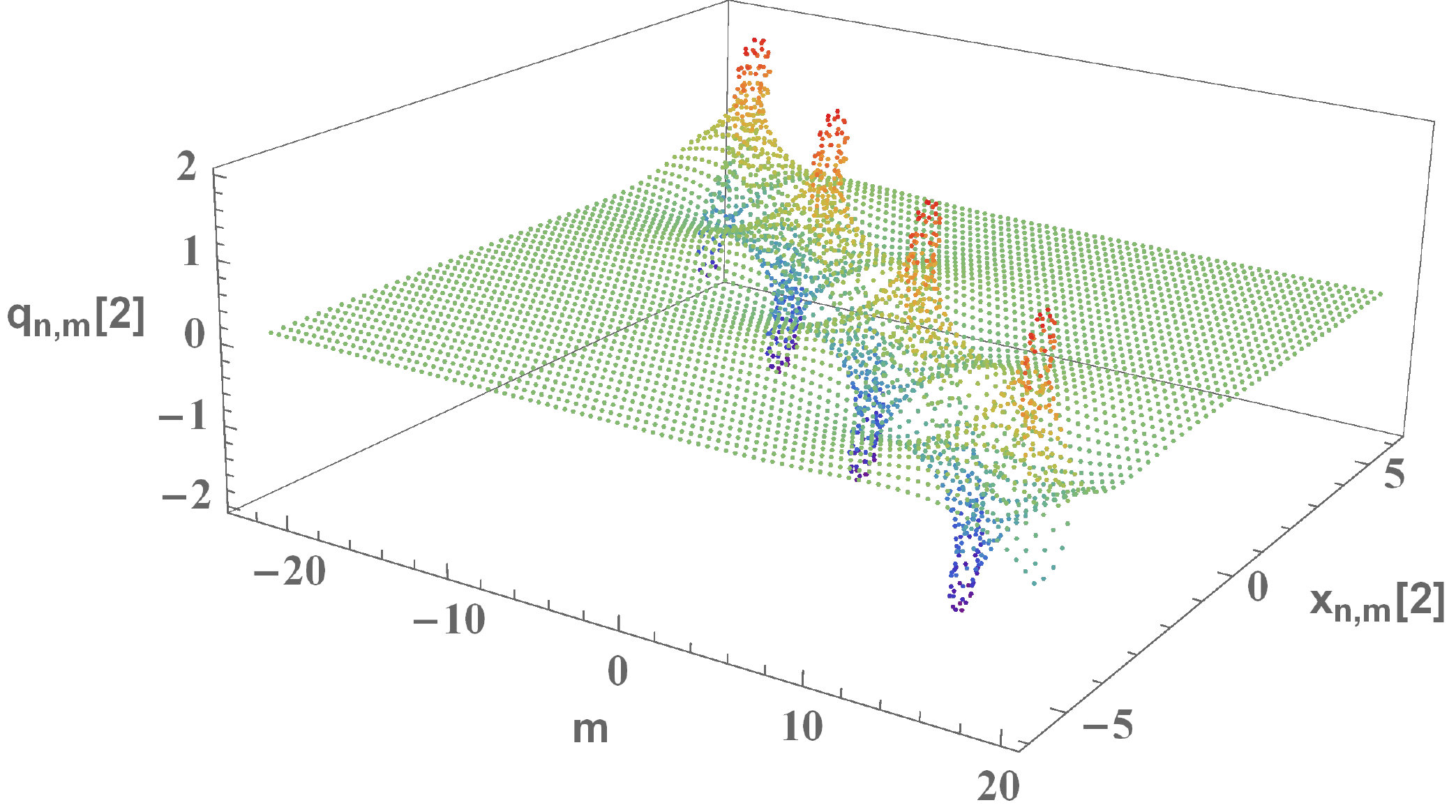}
	\caption{Breather solutions for fully discrete SP equation (\ref{DSP1a})-(\ref{DSP2b}).}
	\label{Breather}
\end{figure*}

Breather solutions are bound state solutions of soliton and
anti-soliton under the specific parametric domain. In such a case,
attractive loop-antiloop pair form a bound structure and oscillate
about each other irrespective of their original shapes. For breather
solution we take $A_{2l}=A_{2l-1}=1$ and $B_{2l-1}=-B_{2l}=1$ and
$\lambda _{1}=-\lambda_2$, $\lambda _{3}=-\lambda_4=$ and
$\lambda_3=\lambda_1^{\ast}$, thus equation (\ref{2foldEXPGen})
become

\begin{equation}
	x_{n,m}[2]= an+\frac{\xi^{(a)}_{n,m}}{\xi^{(c)}_{n,m}}, \quad \quad
	\quad \quad \quad \quad \quad q_{n,m}[2]
	=-\frac{\xi^{(b)}_{n,m}}{\xi^{(c)}_{n,m}}, \label{breatherSOL} \\
\end{equation}
where
\scriptsize
\begin{eqnarray*}
	\xi^{(a)}_{n,m}&=& 2\mbox{i}cd\left(4-\frac{b}{c+\mbox{i} d}\right)^{2 m} (1-a (c+\mbox{i} d))^{2 n}\\
	&&\left(c
	\left(4+\frac{b}{c-\mbox{i} d}\right)^{2 m} (c a-\mbox{i} d a+1)^{2 n}-\mbox{i} d \left(4-\frac{b}{c-\mbox{i}
		d}\right)^{2 m} (-c a+\mbox{i} d a+1)^{2 n}\right)+ \\
	&&2\mbox{i}cd\left(4+\frac{b}{c+\mbox{i} d}\right)^{2 m} (c
	a+\mbox{i} d a+1)^{2 n} \\
	&&\left(\mbox{i} d \left(4+\frac{b}{c-\mbox{i} d}\right)^{2 m} (c a-\mbox{i} d a+1)^{2 n}-c
	\left(4-\frac{b}{c-\mbox{i} d}\right)^{2 m} (-c a+\mbox{i} d a+1)^{2
		n}\right),\\
	\xi^{(b)}_{n,m}&=&  \mbox{i} d 2^{4 m+1}c (c-\mbox{i} d)\left(16+\frac{b^2}{(d-\mbox{i} c)^2}\right)^m
	\left(1-a^2 (c+\mbox{i}
	d)^2\right)^n \\
	&&
	\left(\left(1-\frac{b}{4 c-4 \mbox{i} d}\right)^{2 m} (-c a+\mbox{i} d a+1)^{2 n}+\left(1+\frac{b}{4 c-4 \mbox{i} d}\right)^{2
		m} (c a-\mbox{i} d a+1)^{2 n}\right)- \\
	&&\mbox{i} d 2^{4 m+1}c (c+\mbox{i} d) \left(16+\frac{b^2}{(d+\mbox{i} c)^2}\right)^m \left(1-a^2 (c-\mbox{i}
	d)^2\right)^n \\
	&& \left(\left(1-\frac{b}{4 c+4 \mbox{i} d}\right)^{2 m} (1-a (c+\mbox{i} d))^{2 n}+\left(1+\frac{b}{4 c+4 \mbox{i}
		d}\right)^{2 m} (c a+\mbox{i} d a+1)^{2 n}\right), \\
	\xi^{(c)}_{n,m}&=& 2 \left(c^2+d^2\right)^2
	\left(16+\frac{b^2}{(d-\mbox{i} c)^2}\right)^m \left(16+\frac{b^2}{(d+\mbox{i}
		c)^2}\right)^m
	\left(1-a^2 (c-\mbox{i} d)^2\right)^n \left(1-a^2 (c+\mbox{i} d)^2\right)^n- \\
	&&c^2\left(c^2+d^2\right) \left(4+\frac{b}{c-\mbox{i}
		d}\right)^{2 m}\left(4-\frac{b}{c+\mbox{i} d}\right)^{2 m} (1-a (c+\mbox{i} d))^{2 n} (c a-\mbox{i} d
	a+1)^{2 n}- \\
	&&c^2\left(c^2+d^2\right) \left(4-\frac{b}{c-\mbox{i} d}\right)^{2 m} \left(4+\frac{b}{c+\mbox{i} d}\right)^{2
		m} (-c a+\mbox{i} d a+1)^{2 n} (c a+\mbox{i} d a+1)^{2 n}+\\
	&&d^2\left(c^2+d^2\right) \left(4-\frac{b}{c-\mbox{i} d}\right)^{2 m}
	\left(4-\frac{b}{c+\mbox{i} d}\right)^{2 m} (1-a (c+\mbox{i} d))^{2 n} (-c a+\mbox{i} d a+1)^{2 n}+ \\
	&&d^2\left(c^2+d^2\right)
	\left(4+\frac{b}{c-\mbox{i} d}\right)^{2 m} \left(4+\frac{b}{c+\mbox{i} d}\right)^{2 m} (c a-\mbox{i} d
	a+1)^{2 n} (c a+\mbox{i} d a+1)^{2 n}.
\end{eqnarray*}
\normalsize
First-order breather solution (\ref{breatherSOL}) is shown in Fig.
(\ref{Breather}) for $a=b=c=d=0.5$.

\section{Concluding remarks}
A generalized fully discrete SP equation is explored as an
integrability condition of a linear pair of difference equations. By
considering two continuum limits, generic case has been shown to
reduce to space semi-discrete and time discretized SP equations.
Darboux transformation is applied to the contemporary discrete
linear system. Furthermore, the covariance of the discrete linear
system under Darboux transformation, nontrivial multi-soliton
solutions are calculated in terms of ratios of determinants.
Discrete loop solutions are obtained and the dynamics of repulsive
as well as attractive interactions between loop-loop and
loop-antiloop solutions are discussed. These discrete solutions
under continuum limits were reduced to their respective continuous
counterpart results that existed in literature. Loop-antiloop pairs
also fused into one another to form breather solutions that
oscillate in time irrespective of their original shape. Moreover,
solutions established in this article may have applications in
physics and engineering as the SP equation is regarded as the
physical model for the ultra-short pulse propagations in a nonlinear
medium.  The discretization procedure used in this article can also
be extended to have discrete versions of other nonlinear integrable
systems and analytical solutions of the systems can also be
considered.
	
%	\section*{Acknowledgments}
%	This was was supported in part by......
	
	%Bibliography
	\bibliographystyle{unsrt}  
	\bibliography{references}

\begin{thebibliography}{10}

\bibitem{rothenberg1992space}
Joshua~E Rothenberg.
\newblock Space--time focusing: breakdown of the slowly varying envelope
  approximation in the self-focusing of femtosecond pulses.
\newblock {\em Optics Letters}, 17(19):1340--1342, 1992.

\bibitem{schafer2004propagation}
T~Sch{\"a}fer and CE~Wayne.
\newblock Propagation of ultra-short optical pulses in cubic nonlinear media.
\newblock {\em Physica D: Nonlinear Phenomena}, 196(1-2):90--105, 2004.

\bibitem{rabelo1989equations}
Mauro~L Rabelo.
\newblock On equations which describe pseudospherical surfaces.
\newblock {\em Studies in Applied Mathematics}, 81(3):221--248, 1989.

\bibitem{sakovich2005short}
Anton Sakovich and Sergei Sakovich.
\newblock The short pulse equation is integrable.
\newblock {\em Journal of the Physical Society of Japan}, 74(1):239--241, 2005.

\bibitem{brunelli2006bi}
Jose~Carlos Brunelli.
\newblock The bi-hamiltonian structure of the short pulse equation.
\newblock {\em Physics Letters A}, 353(6):475--478, 2006.

\bibitem{zhang2015conservation}
Zhi-Yong Zhang and Yu-Fu Chen.
\newblock Conservation laws of the generalized short pulse equation.
\newblock {\em Chinese Physics B}, 24(2):020201, 2015.

\bibitem{matsuno2008periodic}
Yoshimasa Matsuno.
\newblock Periodic solutions of the short pulse model equation.
\newblock {\em Journal of mathematical physics}, 49(7):073508, 2008.

\bibitem{matsuno2007multiloop}
Yoshimasa Matsuno.
\newblock Multiloop soliton and multibreather solutions of the short pulse
  model equation.
\newblock {\em Journal of the Physical Society of Japan}, 76(8):084003--084003,
  2007.

\bibitem{parkes2008some}
EJ~Parkes.
\newblock Some periodic and solitary travelling-wave solutions of the
  short-pulse equation.
\newblock {\em Chaos, Solitons \& Fractals}, 38(1):154--159, 2008.

\bibitem{tb2007two}
Bouetou TB and Kofane TC.
\newblock On two-loop soliton solution of the sch{\"a}fer--wayne short-pulse
  equation using hirota’s method and hodnett--moloney approach.
\newblock {\em Journal of the Physical Society of Japan}, 76(2):024004--024004,
  2007.

\bibitem{sakovich2006solitary}
Anton Sakovich and Sergei Sakovich.
\newblock Solitary wave solutions of the short pulse equation.
\newblock {\em Journal of Physics A: Mathematical and General}, 39(22):L361,
  2006.

\bibitem{brunelli2005short}
Jose~Carlos Brunelli.
\newblock The short pulse hierarchy.
\newblock {\em Journal of mathematical physics}, 46(12):123507, 2005.

\bibitem{zhaqilao2017}
Zhaqilao.
\newblock The interaction solitons for the complex short pulse equation.
\newblock {\em Communications in Nonlinear Science and Numerical Simulation},
  47:379--393, 2017.

\bibitem{kingston1982reciprocal}
JG~Kingston and C~Rogers.
\newblock Reciprocal b{\"a}cklund transformations of conservation laws.
\newblock {\em Physics Letters A}, 92(6):261--264, 1982.

\bibitem{kosmann2004discrete}
B~Grammaticos~Y Kosmann-Schwarzbach and T~Tamizhmani.
\newblock Discrete integrable systems.
\newblock {\em Lect. Notes in Physics}, 644, 2004.

\bibitem{davydov1973theory}
Alexander~S Davydov.
\newblock The theory of contraction of proteins under their excitation.
\newblock {\em Journal of Theoretical Biology}, 38(3):559--569, 1973.

\bibitem{kenkre1986self}
VM~Kenkre and DK~Campbell.
\newblock Self-trapping on a dimer: time-dependent solutions of a discrete
  nonlinear schr{\"o}dinger equation.
\newblock {\em Physical Review B}, 34(7):4959, 1986.

\bibitem{papanicolaou1987complete}
N~Papanicolaou.
\newblock Complete integrability for a discrete heisenberg chain.
\newblock {\em Journal of Physics A: Mathematical and General}, 20(12):3637,
  1987.

\bibitem{ablowitz1975nonlinear}
Mark~J Ablowitz and John~F Ladik.
\newblock Nonlinear differential- difference equations.
\newblock {\em Journal of Mathematical Physics}, 16(3):598--603, 1975.

\bibitem{ablowitz1977solution}
Mark~J Ablowitz and John~F Ladik.
\newblock On the solution of a class of nonlinear partial difference equations.
\newblock {\em Studies in Applied Mathematics}, 57(1):1--12, 1977.

\bibitem{hirota1977nonlinear1}
Ryogo Hirota.
\newblock Nonlinear partial difference equations. i. a difference analogue of
  the korteweg-de vries equation.
\newblock {\em Journal of the Physical Society of Japan}, 43(4):1424--1433,
  1977.

\bibitem{hirota1977nonlinear2}
Ryogo Hirota.
\newblock Nonlinear partial difference equations ii; discrete sine-gordon
  equation.
\newblock {\em Journal of the Physical Society of Japan}, 43(6):2074--2078,
  1977.

\bibitem{hirota1977nonlinear3}
Ryogo Hirota.
\newblock Nonlinear partial difference equations iii; discrete sine-gordon
  equation.
\newblock {\em Journal of the Physical Society of Japan}, 43(6):2079--2086,
  1977.

\bibitem{hirota1978nonlinear4}
Ryogo Hirota.
\newblock Nonlinear partial difference equations. iv. b{\"a}cklund
  transformation for the discrete-time toda equation.
\newblock {\em Journal of the Physical Society of Japan}, 45(1):321--332, 1978.

\bibitem{hirota1979nonlinear5}
Ryogo Hirota.
\newblock Nonlinear partial difference equations. v. nonlinear equations
  reducible to linear equations.
\newblock {\em Journal of the Physical Society of Japan}, 46(1):312--319, 1979.

\bibitem{feng2010integrable}
Bao-Feng Feng, Ken-ichi Maruno, and Yasuhiro Ohta.
\newblock Integrable discretizations of the short pulse equation.
\newblock {\em Journal of Physics A: Mathematical and Theoretical},
  43(8):085203, 2010.

\bibitem{feng2015integrable}
Bao-Feng Feng, Ken-ichi Maruno, and Yasuhiro Ohta.
\newblock Integrable semi-discretization of a multi-component short pulse
  equation.
\newblock {\em Journal of Mathematical Physics}, 56(4):043502, 2015.

\bibitem{feng2011discrete}
Bao-Feng Feng, Jun-ichi Inoguchi, Kenji Kajiwara, Ken-ichi Maruno, and Yasuhiro
  Ohta.
\newblock Discrete integrable systems and hodograph transformations arising
  from motions of discrete plane curves.
\newblock {\em Journal of Physics A: Mathematical and Theoretical},
  44(39):395201, 2011.

\bibitem{feng2015integrable1}
Bao-Feng Feng, Junchao Chen, Yong Chen, Ken-ichi Maruno, and Yasuhiro Ohta.
\newblock Integrable discretizations and self-adaptive moving mesh method for a
  coupled short pulse equation.
\newblock {\em Journal of Physics A: Mathematical and Theoretical},
  48(38):385202, 2015.

\bibitem{hanif2020pt}
Y~Hanif, H~Sarfraz, and U~Saleem.
\newblock $\mathcal{PT}$-symmetric semi-discrete short pulse equation.
\newblock {\em Results in Physics}, 19:103522, 2020.

\bibitem{matveev1991darboux}
Vladimir~B Matveev and Salle.
\newblock Darboux transformations and solitons.
\newblock 1991.

\end{thebibliography}

\end{document}